\documentstyle[prb,aps,epsf]{revtex}
\input epsf
\begin{document}
% \draft command makes pacs numbers print
\draft
\twocolumn[
\hsize\textwidth\columnwidth\hsize\csname @twocolumnfalse\endcsname
% \twocolumn

\title{Nucleation of superconductivity in mesoscopic star-shaped superconductors}
\author{D.~A.~Dikin\cite{A1}, V.~Chandrasekhar}
\address{Department of Physics and Astronomy, Northwestern 
University, Evanston, IL 60208, USA \\ }
\author{V.~R.~Misko\cite{A2}, V.~M.~Fomin\cite{A3}, J.~T.~Devreese\cite{A4} }
\address{Theoretische Fysica van de Vaste Stoffen, Universiteit 
Antwerpen (UIA), B-2610 Antwerpen,
Belgium \\ }
\date{\today}
\maketitle

\begin{abstract}
We study the phase transition 
of a star-shaped superconductor, which covers smoothly
the range from  zero to two dimensions with respect to the 
superconducting coherence length $\xi (T)$. Detailed measurements
and numerical calculations show that the nucleation of 
superconductivity in this device is very inhomogeneous, resulting in
rich structure in the superconducting transition as a function of 
temperature and magnetic field. The superconducting order
parameter is strongly enhanced and mostly robust in regions close to 
multiple boundaries.
\end{abstract}

\pacs{PACS numbers: 73.23.-b,74.25.Fy,74.50.+r,74.80.Fp}

]

Many 
characteristics of a superconducting sample are strongly
influenced by its size with respect to fundamental superconducting 
length scales. Experimental and theoretical investigations
over the past few decades  have clearly delineated the differences in 
behavior between one-dimensional,  two-dimensional and
``bulk'' three-dimensional samples in properties such as the critical 
field, critical temperature and  critical current.
Recently \cite{Moshchalkov}, there has  been renewed interest in the 
effect of sample dimension on the properties of
superconductors, fueled in part by their potential use in future 
nanometer scale devices. The complex geometry of such
devices implies that one needs to consider the nucleation of 
superconductivity over a range of length scales in a single
device.

The nucleation of superconductivity in the presence of a magnetic 
field can be investigated by solving the
Ginzburg-Landau(GL) equations for the superconducting order parameter 
$\psi$ and the vector potential $\bf A$ of the magnetic
field $\bf B = \nabla\times A$ \cite{Ginzburg,de Gennes,Tinkham}:

\begin{eqnarray}& &\frac{1}{2m}\left( -i\hbar {\bf\nabla } - 
\frac{2e}{c}{\bf A}\right)^{2}\psi + \alpha \psi + \beta \left|
\psi\right|^{2}\psi = 0,
%\nonumber
\label{glepsi} \\& &{\bf\nabla }^2 {\bf A} = \frac{4\pi i e\hbar 
}{mc}\left( \psi ^{\ast }{\bf \nabla }\psi -\psi {\bf\nabla }
\psi ^{\ast }\right) + \frac{16\pi e^{2}}{mc^{2}}{\bf A}\left| \psi 
\right| ^{2}
%\nonumber
\label{glea}
\end{eqnarray} subject to the boundary condition
\begin{equation}{\bf n}\cdot\lgroup -i\hbar{\bf\nabla} - \frac{2e}{c} 
{\bf A} \rgroup \psi = 0.\label{bc}
\end{equation} Here $\alpha$ and $\beta$ is the GL parameters, and 
$\bf n$ is the unit vector normal to the surface.

Close to the critical temperature $T_c$, $\left| \psi \right| ^{2}$ 
is negligible, and one can analytically solve the
linearized GL equation (\ref{glepsi}) for some simple geometries. 
Saint-James and de Gennes\cite{Saint-James} solved this
linear equation for the case of a magnetic field parallel to the 
surface of a superconductor. They found that
superconductivity could exist at a field  $H_{c3}$ (the so-called 
surface critical field) larger than the upper critical
field $H_{c2}$ in the bulk superconductor, $H_{c3} = 1.69 H_{c2}$. 
More recently\cite{Fomin,Brosens,Klimin},  attention has
focused on superconducting wedges subtending an angle $\gamma$, with 
the magnetic field applied parallel to  the wedge's
edge. In the limit of small $\gamma$, it has been 
predicted\cite{Houghton,VanGelder} that the surface critical field 
can be
greatly enhanced, $H_{c3} = (1.73/\gamma )H_{c2}$, although this has 
never been demonstrated experimentally.  This large
enhancement of $H_{c3}$ can be ultimately traced to the increase in 
the ``energy'' ($-\alpha$) in the GL equation
(\ref{glepsi}) associated with the confinement of the order parameter.

For more complicated geometries, and for regimes further away from 
the superconducting transition, it is necessary to solve
the full GL equations numerically. This has been done for many 
different geometries, including wedges,
rings\cite{Schweigert1,Schweigert2}, square loops\cite{Fomin2,Fomin3} 
and bridges\cite{bridge}. Here we are interested in a
four-pointed ``star'' geometry, shown schematically in Fig. 1. The 
remarkable feature of this geometry is that it encompasses
a wide range of length scales. As we shall see, this results in 
appearance of regions of the sample with characteristically
different behavior in the superconducting regime.

%Fig.~1.
\begin{figure}
%\protect\centerline{\epsfbox{Starprbf1.eps scaled 600}}
%\protect\centerline{\epsfbox{Starprbf1.eps}}
\protect\centerline{\epsfbox{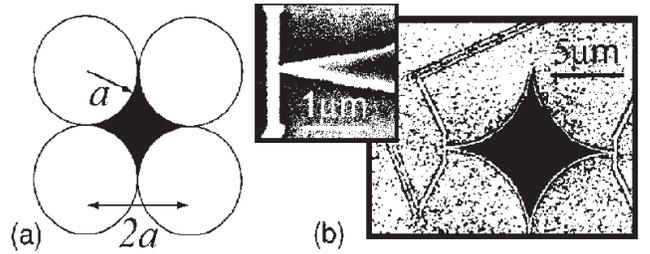}}
\smallskip 
\caption{
(a) Schematic representation of the method that was used to design samples. 
(b) Scanning electron micrograph of one of the samples. The inset shows 
magnified area of the apex.
}
\label{Fig.1}
\end{figure}

The experimental sample is prepared by removing portions of a circle 
of radius $a$ from the corners of  a square of side
$2a$, as shown in Fig.~1 (a). The experimental realization of this 
geometry is shown in Fig.~1(b). The sample is fabricated
by conventional electron beam lithography with a 60 nm thick Al film. 
The apex-to-apex distance is 12 $\mu$m, the minimum
dimension at each apex is less than 100 nm. These dimensions are also 
used in all numerical calculations.  In order to enable
four terminal electrical measurements on the device, narrow 
electrical contacts of the same material are attached to opposite
apices.

%Fig.~2.
\begin{figure}
\protect\centerline{\epsfbox{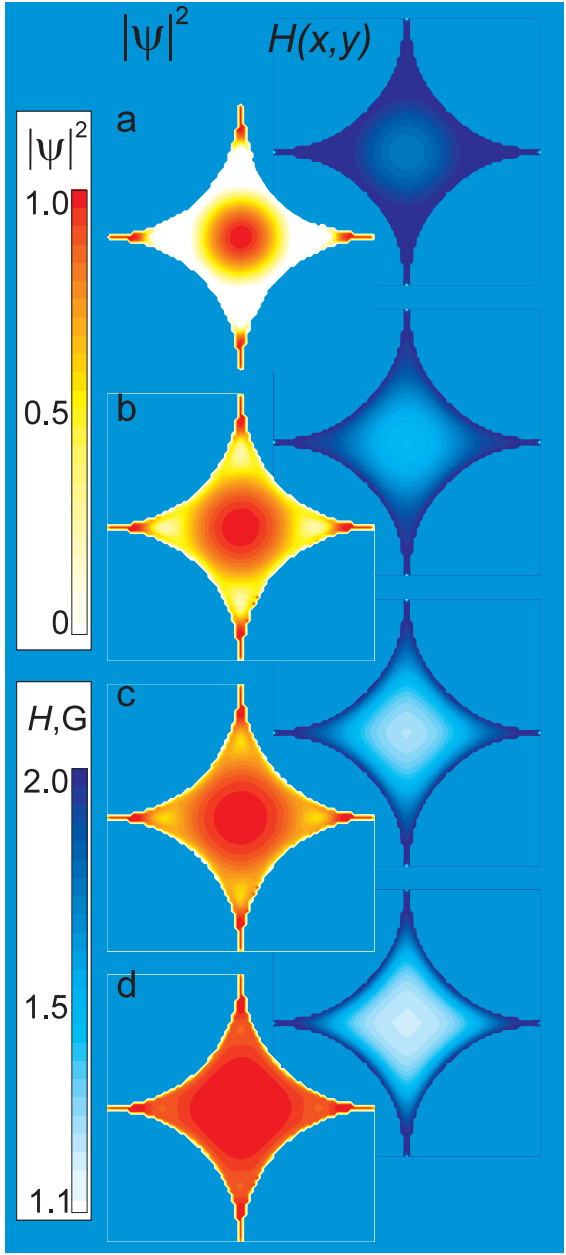}}
\smallskip 
\caption{
Evolution of $\left| \psi (x,y) \right| ^{2}$  and magnetic field 
distribution in the star with temperature for an applied
external field of $H_0 = 0.02H_c(0)$ = 2 G, where $H_c(0)$ = 
100 G is the bulk critical field at zero temperature, at
different temperatures: a - $T/T_c$ = 0.99, b - 0.98, c - 0.97, d - 
0.94. $\left| \psi (x,y)\right| ^{2}$ is normalized to its
maximum value at these temperatures.
}
\label{Fig.2}
\end{figure}

To obtain the solutions of the GL equations (\ref{glepsi}), 
(\ref{glea}) with the boundary conditions (\ref{bc}), we use the
finite-difference method applied earlier for the description of 
superconductivity in a mesoscopic square
loop\cite{Fomin2,Fomin3}, using the dimensions and parameters of the 
experimental sample of Fig. 1. One additional
complication of this geometry for the finite-difference method is the 
infinite sharpness of the apices of the star, which
makes the generation of a mesh for the problem difficult. 
Fortunately, the vertices of the experimental sample have finite
curvature, which results in a definite cut-off length of around 100 
nm for the tip of each apex, as shown in the small inset
to Fig.~1. In our calculations of the order parameter and magnetic 
field distributions, a square mesh has been used with the
density of 800 nodes per side of the sample. At such a high density 
of nodes, the results of calculations are independent of
the mesh.
%Fig.~2.

The solutions of the GL equations with appropriate boundary 
conditions permit us to obtain the spatial distribution of the
order parameter and the magnetic field as a function of temperature 
$T$, as shown in Fig. 2.  As temperature is decreased
slightly from $T_c \,\ (T/T_c = 0.99)$, one observes nucleation of 
superconductivity as evidenced by a non-zero $\left|
\psi(x,y) \right| ^{2}$ in two distinct regions of the sample. The 
first region is a large area in the center, where one
might intuitively expect superconductivity to nucleate. The second 
region is at the apices of the star, where
superconductivity is enhanced by the close proximity of two 
boundaries, and resembles the situation that occurs in a
superconducting ``wedge.'' In between these two regions, $\left| \psi 
\right| ^{2}$ is essentially 0, so that the
superconducting phases of the sample are separated by regions of 
normal phase. The plot of the field distribution shows that
the two superconducting regions are characterized by different 
behavior in a magnetic field: the central region shows the
beginning of Meissner expulsion of the external field, while the 
magnetic field near the boundaries is very close to the
external field value. As the temperature is lowered further, the 
superconducting areas in the two regions grow as expected to
cover almost the entire sample, and the normal regions between the 
apices and the center disappear. At the lowest temperature
shown, the value of $\left| \psi \right| ^{2}$ at the apices and in 
the central region achieves its maximum value
corresponding to this temperature, $\left| \psi (x,y) \right| 
^{2}_{max} \sim (1 - T/T_c) = 0.06$.  The magnetic field at the
center is greatly reduced, showing the presence of a strong Meissner 
effect, but the magnetic field in a narrow region at the
boundaries of the sample is still very close to the external field 
value. It should be noted that the value of $\left| \psi
\right| ^{2}$ at the apices is a maximum, even though the magnetic 
field is also a maximum. In fact, the value of $\left|
\psi \right|^{2}$ at the apices does not change appreciably over the 
temperature range shown in Fig.~2, demonstrating vividly
the fact that superconductivity is the most robust in regions close 
to multiple boundaries.
%Fig.~3.

The electrical contacts in the sample are placed at opposite apices, 
so that the electrical current must traverse these
apices, the central region, as well as any normal regions in between. 
Consequently, the inhomogeneous nature of
superconductivity nucleation as the sample is cooled through its 
transition imparts a distinctive shape 

%Fig.~3.
\begin{figure}
\protect\centerline{\epsfbox{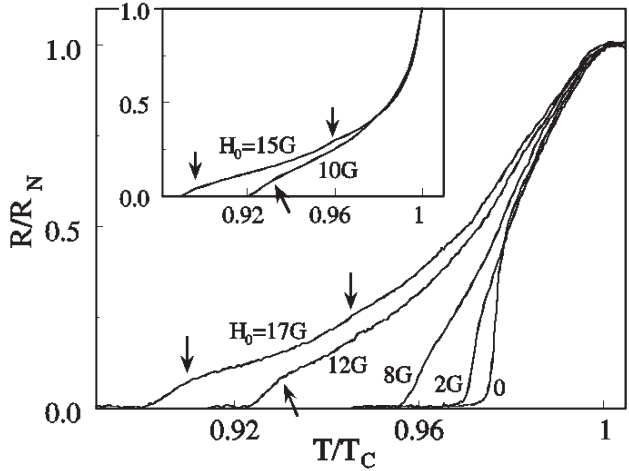}}
\smallskip 
\caption{
Experimentally measured resistive transition, taken at 
different magnetic
fields. The inset shows the numerically calculated $R(T)$. $T_c$ = 
1.32 K, and $R_N$ = 12 $\Omega$.
}
\label{Fig.3}
\end{figure}

\noindent
to $R(T)$, as shown
in Fig.~3. The normal-to-superconducting transition starts with a 
decrease in resistance over a relatively broad range in
temperature even at $H_0 = 0$. In this temperature range, 
superconductivity begins to nucleate at the apices and at the
central part, as was shown in Fig.~2, but the regions near the 
apices, being narrower, contribute a large fraction to the
resistance change. As the temperature is reduced further, there is a 
much sharper drop in the resistance when $R$ is a little
less than half the normal state resistance $R_N$. Since the central 
region and the apices are already superconducting, this
rapid change corresponds to the normal-to-superconducting transition 
of the ``necks'' between the apices and the central
region. Calculations of the  superconducting transition confirm this picture.

Numerically, we calculate the resistance of the sample using the 
order parameter distribution. As the order parameter and
magnetic field distributions, the resistance of the sample is 
calculated on a rectangular mesh. The resistance of a mesh cell
in the sample is determined by the value of $\left| \psi (x,y) 
\right| ^{2}$ in the cell, being zero if $\left| \psi (x,y)
\right| ^{2} > 0$, and equal to the normal state resistance if 
$\left| \psi (x,y) \right| ^{2} = 0$. One must also take into
account the Josephson coupling between the superconducting regions, 
which reduces the resistance of the ``necks''. We suppose
that a region of the normal metal of length $\leq p\cdot \xi$ (where 
$p \approx 2$ is an adjustable parameter) between
superconducting ``islands'' does not contribute to the resistance in 
the circuit. The inset of Fig.~3 shows result of this
calculation for two different field values. The theoretical curves 
reproduce the main qualitative features of the
experimental curves, including the ``kinks'' (denoted by the arrows) 
at lower resistance, and are in reasonably good
quantitative agreement as well. The calculated position of the 
beginning of the resistive transition compares rather well
with the experimental curves when the tunneling of the 
superconducting electrons through the normal metal is taken into
consideration.
%Fig.~4.

The behavior of the sample in a magnetic field (Fig. 4) exhibits even 
more distinctly the difference between the nature of
the superconductivity in the center and the apices. While $\left| 
\psi (x,y) \right| ^{2}$ in the central region is rapidly
attenuated with increasing magnetic field, its value in the apices 
does not show a large change. At large magnetic fields,
only the apices are superconducting. This would imply that the 
contribution to the resistance of the sample from the apices
is not strongly affected by magnetic field. Evidence for this can be 
seen directly in the experimental transition curves. The
resistance of the sample at the top of the transition, where the 
decrease in resistance is primarily due to the nucleation of
superconductivity in the apices, shows only a weak dependence on 
magnetic field. At lower values of resistance, where one has
contributions from the central region as well as the ``necks'', the 
field dependence is much stronger, reflecting the
two-dimensional nature of the superconductivity in these regions.
%Fig.~5.

Even more striking is the magnetic phase diagram ($T_c$ vs. $H$) of 
the sample. Typically, this is obtained by varying the
temperature to maintain the resistance of the sample at the midpoint 
of the transition ($ R/R_N = 0.5$) while changing the
magnetic field. Since our sample is inhomogeneous, and it is not 
clear how one would define the unique $T_c$, we have
measured the phase diagram with the sample biased at various points 
$R/R_N$ of the superconducting transition. Figure 5 shows
the result of these measurements. Although  the curves are different, 
each one shows two distinct regions, separated by a
well-defined ``kink'' denoted by the arrow. At low fields, the curves 
show a quasi-linear dependence on $H_0$, which turns
into a quadratic dependence at higher magnetic fields. It is well 
known that the critical temperature of a superconductor
varies linearly with the magnetic field in two dimensions, and 
quadratically in one dimension\cite{Tinkham}. The ``kink'' in
each curve defines the crossover from two-dimensional behavior at low 
fields to one-dimensional behavior at higher fields.
This behavior is in agreement with the evolution of the 
superconducting regions of the sample shown in the numerical
simulations of Fig. 3. At low fields, the central region and the 
``necks'' of the sample are superconducting, giving a
two-dimensional characteristic to the phase diagram. At higher 
fields, superconductivity in these regions is attenuated,
leaving a finite order parameter only in the apices, which gives rise 
to a quadratic dependence on the magnetic field. It is
interesting to note that the low-field linear behavior of all the 
curves extrapolates to a single temperature at $H_0= 0$,
while the high-field quadratic behavior of all the curves also 
extrapolates to a single (but different) temperature. The
different temperatures clearly point out the inhomogeneous nature of 
the superconducting transition in this device.

\twocolumn[
\hsize\textwidth\columnwidth\hsize\csname @twocolumnfalse\endcsname

%Fig.~4.
\begin{figure}
\protect\centerline{\epsfbox{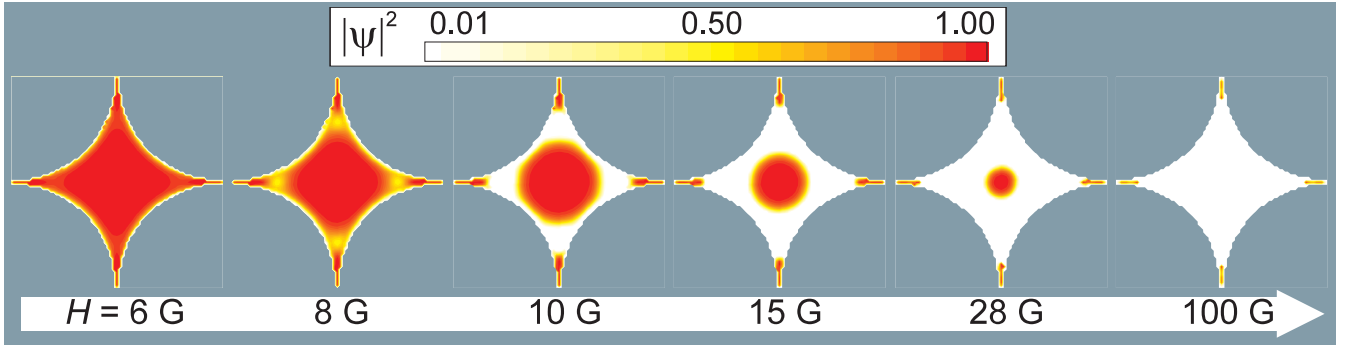}}
\smallskip 
\caption{
Evolution of $\left| \psi (x,y) \right| ^{2}$ with magnetic field at 
$T/T_c$ = 0.8. $\left| \psi (x,y) \right| ^{2}$ is
normalized to its maximum value at this temperature.
}
\label{Fig.4}
\end{figure}

]

\noindent

%Fig.~5.
\begin{figure}
\protect\centerline{\epsfbox{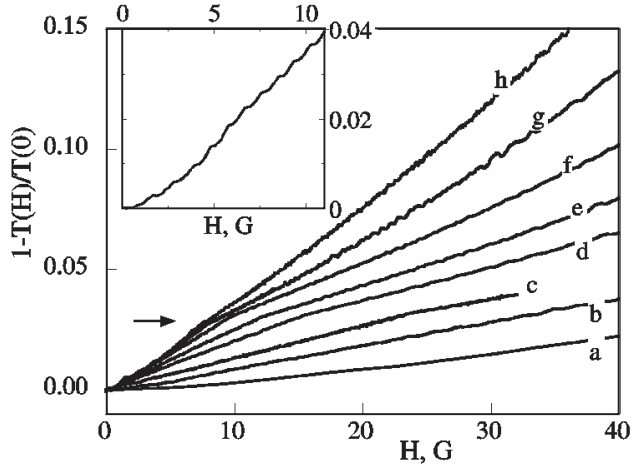}}
\smallskip 
\caption{
$T-H$ phase diagrams measured at $R/R_N$ = 0.58 (a), 0.42 
(b), 0.33 (c), 0.25 (d), 0.21 (e), 0.17 (f), 0.125 (g), 0.08
(h). Inset shows the Little-Parks\cite{Little} oscillations at 
smaller field for $R/R_N$ = 0.04.
}
\label{Fig.5}
\end{figure}

In conclusion, we have analyzed the resistive transition in a 
star-shaped sample, which combines properties of
two-dimensional and zero-dimensional superconductors in a unique 
manner. Occurrence of qualitatively different regions in the
resistive transition reflects a smooth change from 2D to 0D 
superconducting regime when increasing applied magnetic field or
temperature.

\smallskip

We thank V. V. Moshchalkov for valuable discussions. This 
work was supported at Northwestern University by the
National Science Foundation through grant DMR-0201530 and by the 
David and Lucile Packard Foundation, and at Universiteit
Antwerpen by the IUAP, the GOA BOF UA, the FWO-V, WOG (Belgium) and 
the ESF Programme VORTEX.

\end{document}